\documentclass[prd,nofootinbib,english]{revtex4}

\usepackage{graphicx,float}
\usepackage{amsmath,amssymb,amsfonts}
\usepackage{epsfig,color}
\usepackage[thinlines]{easytable}
\usepackage{pdfpages}
\usepackage{array}
\usepackage{cancel}
\usepackage{mathtools}
\usepackage{accents}
\usepackage{subfigure}
\usepackage{enumitem}
\usepackage[dvipsnames]{xcolor}
 \usepackage{hyperref}
\hypersetup{
    colorlinks=true,
    linkcolor=blue,
    filecolor=magenta,      
   citecolor=blue
}
\usepackage{makecell}
\usepackage{multirow}

\begin{document}

\title{Linear perturbations of symmetric teleparallel gravity on Minkowski background}

\author{Dehao Zhao}
\email{dhzhao@ucas.ac.cn}
\affiliation{School of Physical Sciences, University of Chinese	Academy of Sciences, Beijing 100049, China}

\begin{abstract}
Symmetric teleparallel gravity (STG) can be regarded as a modified gravity theory that lacks diffeomorphism symmetries, which complicates the calculation of its degrees of freedom. In this study, we analyze the linear perturbations of general STG models on a Minkowski background, considering both scenarios with and without scalar couplings. Furthermore, we provide lower bounds for the number of degrees of freedom associated with each model.
\end{abstract}

\maketitle

\section{Introduction}

Although general relativity (GR) has achieved great success, it faces the difficulties in both theoretically and observationally, such as singularity of the universe, dark matter and dark energy. Therefore, GR is now generally considered as an effective theory and thought to be modified at high energy and/or large length scales. In the framework of metric-affine theory \cite{Hehl:1994ue}, GR is characterized as a constrained metric-affine gravity that necessitates the vanishing of both torsion and nonmetricity tensors. In this context, the Riemann curvature tensor is employed to describe gravitational effects. Additionally, there are two other gravity theories: teleparallel gravity (TG) \cite{Hayashi:1979qx,Linder:2010py,Cai:2015emx}, in which both curvature and nonmetricity tensors vanish while utilizing the torsion tensor to characterize gravity; and symmetric teleparallel gravity (STG) \cite{Nester:1998mp,BeltranJimenez:2017tkd}, where curvature and torsion tensors are absent, relying instead on the nonmetricity tensor to account for gravitational phenomena. This paper will focus on the STG theory.

STG theory takes nonmetricity tensor
\begin{eqnarray}
	Q_{\alpha\mu\nu}=\nabla_\alpha g_{\mu\nu}=\partial_\alpha g_{\mu\nu}-\Gamma^{\lambda}{}_{\alpha\mu}g_{\lambda\nu}-\Gamma^{\lambda}{}_{\alpha\nu}g_{\mu\lambda}
\end{eqnarray}
to account for gravitational phenomena, which takes metric and affine connection as fundamental variables. Since both the curvature and torsion tensors in STG are identically zero, one can always choose a particular coordinate system in which affine connection vanishes. The adoption of this particular coordinate system means one have fixed a gauge known as the "coincident gauge." There exists a special STG theory whose action differs from the Einstein-Hilbert action only by a surface term. Thus, this special STG theory is dynamically equivalent to GR. For this reason, this theory is also called  "symmetric teleparallel equivalent of general relativity" (STEGR). Therefore, based on the STEGR with a Lagrangian density $\sqrt{-g}Q$, one can readily construct modified gravity models that make slight adjustments to General Relativity in order to satisfy various experiment results, such as $f(Q)$ gravity \cite{BeltranJimenez:2019tme,Lazkoz:2019sjl,Zhao:2021zab}. Many modified gravity models exist within the STG framework and have been employed to elucidate various aspects of gravitational phenomena \cite{Hassan:2022hcb,Zhao:2019xmm,Solanki:2022ccf,Hassan:2021egb,Gakis:2019rdd,Conroy:2019ibo,Zhao:2024kri}.

Since it is always possible to choose a coincident gauge where the affine connection vanishes, STG theory can also be regarded as a pure metric gravity theory but lacks diffeomorphism symmetries. Consequently, the number of degrees of freedom (DOFs) in general STG models exceeds two. The standard approach to compute the number of DOFs is through Hamiltonian analysis based on the Dirac-Bergmann algorithm \cite{Dirac:1950pj,Anderson:1951ta}.However, there appear to be challenges in applying Hamiltonian analysis to STG models. For instance, in the $f(Q)$ model, the authors of \cite{Hu:2022anq} argued that there are eight DOFs in four dimensional spacetime using Hamiltonian analysis. But the authors in \cite{DAmbrosio:2023asf} pointed out that f(Q) gravity breaks one of the basic assumptions of the algorithm, i.e., the equations of Lagrangian multipliers become differential equations from linear algebraic equations. Hence the Dirac-Bergmann algorithm fails in $f(Q)$ case.  By analyzing the primary constraints of $f(Q)$ models, they also gave a upper bound of the number of DOFs, which is seven. Latter the authors in \cite{Tomonari:2023wcs} asserted that they have solved the problem presented by \cite{DAmbrosio:2023asf} and said that there are six DOFs. But it is wrong. The Papers \cite{Gomes:2023tur,Rao:2023nip} calculate the linear perturbations of $f(Q)$ model, and showed that there are seven DOFs on cosmological background. And this can be considered as a lower bound of $f(Q)$ model. Thus far, we can confidently conclude that there are seven DOFs associated with the $f(Q)$ model. Similar issue also occurs in TG framework, such as in $f(T)$ models, different calculations of number of DOFs gave different results \cite{Li:2011rn,Ferraro:2018tpu,Blagojevic:2020dyq}. Therefore, to have a first look at the number of DOFs of STG theory, this work primarily focuses on calculating linear perturbations of general STG models both with and without scalar couplings on a Minkowski background. From these findings, we aim to establish lower bounds applicable to numerous STG models. We think this work can improve our understanding of STG theory and also diffeomorphism symmetry breaking gravity theories. 

The contents of this paper are organized as follows. In Section \ref{section2}, we provide a brief introduction to STG theory and outline the primary model considered in this study. Additionally, we present a proof demonstrating that if the metric and affine connection adhere to the symmetries of the background spacetime, then the equations of motion (EOMs) will also conform to these symmetries. In Section \ref{section3}, we conduct linear perturbation analyses of general STG theory on a Minkowski background and establish lower bounds for each case. In Section \ref{section4}, we explore scenarios involving scalar couplings and calculate linear perturbations within a Minkowski background. Finally, Section \ref{section5} presents our conclusions.

\section{Symmetric teleparallel gravity}\label{section2}

\subsection{Action and equations of motion}

In the framework of metric-affine theory, both the metric and affine connection are considered fundamental variables. Utilizing these two essential components, we can define three important tensors: curvature, torsion
\begin{eqnarray}
    {R_{\mu\nu\sigma}}^{\rho}= - \partial_{\mu} {\Gamma^{\rho}}_{\nu\sigma} + \partial_{\nu} {\Gamma^{\rho}}_{\mu\sigma} +{\Gamma^{\lambda}}_{\mu\sigma}{\Gamma^{\rho}}_{\nu\lambda} - {\Gamma^{\lambda}}_{\nu\sigma}{\Gamma^{\rho}}_{\mu\lambda},\quad {T^{\lambda}}_{\mu\nu} = {\Gamma^{\lambda}}_{\mu \nu} - {\Gamma^{\lambda}}_{\nu\mu},
\end{eqnarray}
and nonmetricity tensor
\begin{eqnarray}
	Q_{\alpha\mu\nu}=\nabla_\alpha g_{\mu\nu}=\partial_\alpha g_{\mu\nu}-\Gamma^{\lambda}{}_{\alpha\mu}g_{\lambda\nu}-\Gamma^{\lambda}{}_{\alpha\nu}g_{\mu\lambda},
\end{eqnarray}
where the signature of metric is $\{-1,1,1,1\}$. Using the definitions of torsion and nonmetricity tensor, we can get the general form of affine connection
\begin{eqnarray}
	{\Gamma^{\lambda}}_{\mu \nu}  =  {\mathring{\Gamma}^{\lambda}}{}_{\mu \nu} + {S^{\lambda}}_{\mu\nu},
\end{eqnarray}
where ${\mathring{\Gamma}^{\lambda}}{}_{\mu \nu}$ is the usual Levi-Civita connection
\begin{eqnarray}
	{\mathring{\Gamma}^{\lambda}}{}_{\mu \nu} = \frac{1}{2} g^{\lambda\rho}( \partial_{\mu}g_{\rho\nu} + \partial_{\nu}g_{\mu\rho} - \partial_{\rho}g_{\mu\nu} ),
\end{eqnarray}
and 
\begin{eqnarray}
	S_{\lambda\mu\nu}= -\frac{1}{2}( T_{\mu\nu\lambda} +T_{\nu\mu\lambda} - T_{\lambda\mu\nu} ) -\frac{1}{2}( Q_{\mu\nu\lambda} +Q_{\nu\mu\lambda} - Q_{\lambda\mu\nu} )
\end{eqnarray}
is called the distortion tensor. We denote $\mathring{\nabla}$ as the covariant derivative operator corresponding to the Levi-Civita connection ${\mathring{\Gamma}^{\lambda}}{}_{\mu\nu}$; furthermore, from this point onward, all quantities marked with a ring above them will be associated with this Levi-Civita connection unless otherwise specified. Then different gravity theories select different classes of affine connection to describe gravity effects. For instance, GR chooses the subclass in which both torsion and nonmetricity tensor vanish.
The gravity theory we consider in this paper, symmetric teleparallel gravity (STG) takes another subclass where both curvature and torsion tensor are zero, thereby utilizes the nonmetricity tensor to characterize gravity effects. Since the curvature and torsion tensors are zero in STG, one can always choose a particular coordinate system in which affine connection is zero. Consequently, in arbitrary coordinate systems, the affine connection can be written as
\begin{eqnarray}\label{2.STG affine connection}
    \Gamma^{\lambda}{}_{\mu\nu}=\frac{\partial x^\lambda}{\partial y^\beta} \partial_\mu\partial_\nu y^\beta,
\end{eqnarray}
where $x^\mu$ are coordinates, and $y^\mu$ can be considered as four scalar fields which can also construct coordinate system. From here, we can see that the constrained affine connection in STG can fully determined by $y^\mu$.

The most general action which is quadratic in nonmetricity tensor is
\begin{eqnarray}\label{2.action QQ}
	S=\frac{1}{2}\int d^4x \sqrt{-g}\left(c_1 Q_{\alpha\mu\nu}Q^{\alpha\mu\nu}+\tilde{c}_2Q_{\alpha\mu\nu}Q^{\mu\nu\alpha}+c_3Q^{\alpha\mu}{}_{\mu}Q_{\alpha\nu}{}^{\nu}+c_4Q^{\alpha\mu}{}_{\mu}Q^{\nu}{}_{\nu\alpha}+\tilde{c}_5Q^{\mu}{}_{\mu}{}^{\alpha}Q^{\nu}{}_{\nu\alpha}\right)+S_{m},
\end{eqnarray}
where $c_1$, $\tilde{c}_2$, $c_3$, $c_4$, $\tilde{c}_5$ are all constant and $S_{m}$ is the matter action. The fundamental variables are metric $g_{\mu\nu}$ and four scalar fields $y^\mu$ which construct the STG affine connection (\ref{2.STG affine connection}). Using the relation of curvature tensors correspond to different affine connections, one can find that if the coupling constants satisfy
\begin{eqnarray}\label{2.GR condition}
    c_1=-\frac{1}{4},\quad \tilde{c}_2=\frac{1}{2},\quad c_3=\frac{1}{4},\quad c_4=-\frac{1}{2},\quad \tilde{c}_5=0,
\end{eqnarray}
the action (\ref{2.action QQ}) can be written as 
\begin{eqnarray}\label{2.action R+Q}
    S=\frac{1}{2}\int d^{4}x   \sqrt{-g}\left[  \mathring{R}+\mathring{\nabla}_{\mu}(Q^{\mu} - \tilde{Q}^{\mu}) \right]+S_m,
\end{eqnarray}
where
\begin{eqnarray}
	Q_{\alpha} = g^{\sigma\lambda}Q_{\alpha\sigma\lambda},\quad \tilde{Q}_{\alpha} =  g^{\sigma\lambda}Q_{\sigma\alpha\lambda}
\end{eqnarray}
are two different contractions of nonmetricity tensor. Since the action (\ref{2.action QQ}) under the condition (\ref{2.GR condition}) is merely a total derivative away from the Einstein-Hilbert action, they are dynamically equivalent. For this reason, this model is referred to as the "symmetric teleparallel equivalent of general relativity" (STEGR). Given that GR is highly regarded and aligns with most experimental results, one can readily construct modified gravity models that deviate slightly from GR based on STEGR.

Here we will consider the most general action (\ref{2.action QQ}).  The variation of this action with respect to the metric yields the metric equations of motion (EOMs),
\begin{eqnarray}\label{2.eom}
	2\nabla_{\alpha}P^{\alpha\mu\nu}+Q_{\alpha}P^{\alpha\mu\nu}-\frac{1}{2}Qg^{\mu\nu}+\left(Q^{(\mu|\alpha\beta|}P^{\nu)}{}_{\alpha\beta}+2Q^{\alpha\beta(\mu}P_{\alpha\beta}{}^{\nu)}\right)=\tau^{\mu\nu}.
\end{eqnarray}
where $\tau^{\mu\nu}$ is energy-momentum tensor
\begin{eqnarray}
	\tau^{\mu\nu} = \frac{2}{\sqrt{-g}} \frac{\delta S_{m}}{\delta g_{\mu\nu}},
\end{eqnarray}
and for writing convenient, we have defined
\begin{eqnarray}        
P_{\alpha\mu\nu}=c_1Q_{\alpha\mu\nu}+\tilde{c}_2Q_{(\mu\nu)\alpha}+c_3g_{\mu\nu}Q_{\alpha}+\frac{c_4}{2}(\tilde{Q}_{\alpha}g_{\mu\nu}+g_{\alpha(\mu}Q_{\nu)})+\tilde{c}_{5}g_{\alpha(\mu}\tilde{Q}_{\nu)}.
\end{eqnarray}
As a constraint within metric-affine theory, it is also necessary to derive the EOMs for the affine connection. However, due to diffeomorphism invariance, these EOMs can be obtained from those of the metric  \cite{Zhao:2021zab,Li:2021mdp}. Therefore, we can only consider the EOMs of metric.

\subsection{Background solutions}

From the equations of motion (\ref{2.eom}), we can easily find that 
\begin{eqnarray} \label{2.minkowski background}
	\bar{g}_{\mu\nu}=\eta_{\mu\nu}, \quad \bar{\Gamma}^{\lambda}{}_{\mu\nu}=0
\end{eqnarray}
is a solution for vacuum, where the bar over the head means that they are quantities of background. This paper mainly focus on the perturbations on this background. Then basing on this background, we can define perturbed metric and affine connection as
\begin{eqnarray}
	g_{\mu\nu}=\eta_{\mu\nu}+\delta g_{\mu\nu},\quad \Gamma^{\lambda}{}_{\mu\nu}=\partial_\mu\partial_\nu u^{\lambda},
\end{eqnarray}
where we have used the form of STG affine connection (\ref{2.STG affine connection}) and $y^\mu=\bar{y}^\mu+u^\mu=x^\mu+u^\mu$. Then the perturbed nonmetricity tensor is
\begin{eqnarray}
	Q_{\alpha\mu\nu}&=&\partial_\alpha g_{\mu\nu}-\Gamma^{\lambda}{}_{\alpha\mu}g_{\lambda\nu}-\Gamma^{\lambda}{}_{\alpha\nu}g_{\mu\lambda}\nonumber\\
	&=&\partial_{\alpha}(\delta g_{\mu\nu}-\partial_{\mu}u_{\nu}-\partial_{\nu}u^{\mu}).
\end{eqnarray}
It is easy to see that the nonmetricity tensor is of first order. Consequently, since we are only considering linear perturbations within the framework of STG theory, the action (\ref{2.action QQ}) represents the most general formulation that can be entirely constructed from the nonmetricity tensor and will influence linear perturbations on a Minkowski background. Further, if we define
\begin{eqnarray}\label{2.H=}
	H_{\mu\nu}\equiv \delta g_{\mu\nu}-\partial_{\mu}u_{\nu}-\partial_{\nu}u^{\mu},
\end{eqnarray}
the action of (\ref{2.action QQ}) can be written as
\begin{eqnarray}\label{2.action HH}
	S=\frac{1}{2}\int d^4x \left(c_1\partial_\alpha H_{\mu\nu}\partial^{\alpha}H^{\mu\nu}+c_2\partial_{\alpha}H_{\mu\nu}\partial^{\mu}H^{\nu\alpha}+c_3\partial_\alpha H\partial^\alpha H +c_4\partial_\alpha H\partial_\beta H^{\alpha\beta}\right)
\end{eqnarray}
where we have defined $c_2=\tilde{c}_2+\tilde{c_5}$ and $H=\eta^{\mu\nu}H_{\mu\nu}$. It should be emphasized that all perturbation variables are lowed and uped by background metric $\eta^{\mu\nu}$ and also we define $\partial^{\alpha}=\eta^{\alpha\beta}\partial_\beta$.

It is straightforward to verify that the action (\ref{2.action HH}) represents the most general formulation of the kinetic term for a spin-two field on a Minkowski background. And the choice of coupling constants
\begin{eqnarray}
	c_1=-\frac{1}{4},\quad c_2=\frac{1}{2},\quad c_3=\frac{1}{4},\quad c_4=-\frac{1}{2}
\end{eqnarray}
also corresponds to the kinitic term of Firze-Pauli action, which is a famous gravity model in massive gravity \cite{Hinterbichler:2011tt,deRham:2014zqa,deRham:2010kj}. STG theory shares many of the same features with massive gravity. For instance, if we choose the coincident gauge in which affine connection vanishes, then STG theory reduces to a pure metric gravity theory without diffeomorphism invariance. When considering perturbations on a general background, the final action contains both kinetic term and mass term of metric perturbations and this can be considered as a action of massive gravity. However, there are notable differences between STG and conventional massive gravity models. Firstly, massive gravity models mainly focus on the mass term and the kinetic term of them is just Einstein-Hilbert action; but STG models mainly focus on the kinetic term since the action are constructed by the covariant derivative of metric. Secondly, the GR solutions are also the solutions of the general massive gravity models since the spin-two field are defined as a perturbations on GR background; but in STG models, whether there are still GR solutions needs to check. 

One can see that the quadratic action (\ref{2.action HH}) contains only the kinetic term of a spin-two field. Is it possible to incorporate a mass term on the Minkowski background described in (\ref{2.minkowski background})? A straightforward ides is to add a cosmological constant term to the action (\ref{2.action QQ}). In this case, for vacuum solution, the energy momentum tensor is $\tau_{\mu\nu}=-\Lambda g_{\mu\nu}$, where $\Lambda$ is the cosmological constant. However, it is readily apparent that the Minkowski background (\ref{2.minkowski background}) ceases to be a solution under these conditions.  Therefore, in the framework of STG theory, we can not add a mass term of metric perturbations on Minkowski background. In GR, we know that solutions derived from the Einstein-Hilbert action along with a cosmological constant yield de Sitter or anti-de Sitter (dS/AdS) spacetime. Herein we want to check whether dS/AdS spacetime is still the background solution of action (\ref{2.action QQ}) with a cosmological constant.

For dS spacetime, the metric ansatz in inflation coordinate system can be written as
\begin{eqnarray}
	ds^2=-dt^2+e^{\lambda t}(dx^2+dy^2+dz^2),
\end{eqnarray}
where $\lambda$ is a constant. Given that the STG theory is a metric-affine theory, we assume that the affine connection vanishes in this coordinate system. We define the left-hand-side of Eq.(\ref{2.eom}) as $N^{\mu\nu}$ and incorporate both the metric and affine connection into the equations of motion (\ref{2.eom}). The non-vanishing components of $N^{\mu\nu}$ are
\begin{eqnarray}\label{2.offdiagonal Nxx}
	N^{tt}=\frac{3\lambda^2}{2}(c_1+9c_3+3c_4),\quad N^{xx}=N^{yy}=N^{zz}=-\frac{3\lambda^2}{2}e^{-\lambda t}(c_1+3c_3).
\end{eqnarray}
For GR parameters (\ref{2.GR condition}), we have $N^{\mu\nu}=-(3\lambda^2/4)g^{\mu\nu}$, then the cosmological constant is $3\lambda^2/4$. However, for the STG model described by (\ref{2.action QQ}) with an associated cosmological constant, it ceases to be a solution due to the emergence of non-vanishing off-diagonal terms. Since we have fixed the coincident gauge, different coordinate system are not equivalent. Consequently, we also examined dS/AdS metrics across various coordinate systems; none yielded solutions. For instance, consider AdS spacetime represented in static coordinates as follows
\begin{eqnarray}
	ds^2=-\cosh^2 (\Psi)dt^2+l^2 [d\Psi^2+\sinh^2(\Psi)(d\theta^2+\sin^2(\theta)d\phi^2)],
\end{eqnarray}
in this case, there are also non-vanishing off-diagonal components
\begin{eqnarray}\label{2.offdiagonal Npp}
	N^{\Psi\theta}=N^{\theta\Psi}=\left[2c_1+\tilde{c}_2+2c_3+c_4+(2c_1+\tilde{c}_2+6c_3+3c_4)\cosh(2\Psi)\right]\cot(\theta)\mathrm{csch}^3(\Psi)\mathrm{sech}(\Psi).
\end{eqnarray}
One can see that a typical feature is the presence of non-vanishing off-diagonal terms in the metric equations. The reason for these non-vanishing off-diagonal terms is straightforward. Given that we have fixed the coincident gauge in which the affine connection is set to zero, the STG theory can be regarded as a purely metric theory. Consequently, the gravitational component of the action lacks diffeomorphism invariance. However, generally speaking, the matter sector maintains diffeomorphism invariance, as seen in vacuum scenarios. This symmetry requirement imposes additional constraints and similar features emerge in various modified gravity theories, such as Palatini $f(R)$ formulation \cite{Sotiriou:2006qn,Sotiriou:2008rp}.

To find suitable background solutions within the framework of STG gravity, it is common practice to impose a constraint on the affine connection \cite{Hohmann:2019nat,Hohmann:2021ast}. This constraint requires that the affine connection also adheres to all symmetries of the background spacetime. Mathematically, this condition can be expressed as follows
\begin{eqnarray}
	\mathcal{L}_{\zeta}\Gamma^{\lambda}{}_{\mu\nu}=0,
\end{eqnarray}
where $\mathcal{L}$ denotes the Lie derivative and $\zeta$ represents all Killing vectors associated with the background. We will prove in next subsection that this requirement is equivalent to state that the nonmetricity tensor must satisfy all symmetries of the background spacetime. It is well-known that dS/AdS spacetimes are maximally symmetric spacetimes. From paper \cite{Zhao:2021zab}, we can obtain that if nonmetricity tensor satisfies all the symmetries of  maximally symmetric spacetimes in four dimension, it must be zero. Consequently, it follows that $N^{\mu\nu}$ should also equal zero and it means that this type of affine connection is not a solution. Therefore, employing this method may not yield our desired results. Thus, finding a dS/Ads solution under such circumstances is challenging.
 
 \subsection{A proof that   }
 
 The STG theory considers the metric and constrained affine connection as fundamental variables. The form of the metric is selected to ensure that it satisfies all symmetries of spacetime. In this section, we will demonstrate that if the affine connection also conforms to all symmetries of the background spacetime, then the equations of motion (EOMs) for the gravitational component, denoted as \( E^{\mu\nu} \), will likewise satisfy these symmetries:
 \begin{eqnarray}
 	\mathcal{L}_{\zeta}E^{\mu\nu}=0.
 \end{eqnarray}
 We believe this is why such a choice of affine connection generally yields correct results. As we have not encountered a proof for this assertion in existing literature, we shall provide one in this subsection.
 
Since the affine connection is not a tensor, we need to know what is the meaning of Lie derivative of affine connection. The meaning is as follows: under a diffeomorphism transformation $x^{\mu}\rightarrow x^{\mu}+\zeta^{\mu}$, the transformed affine connection dose not change the parallel structure. Consequently, the transformed parallel vector field is also a parallel vector field. Therefore the Lie derivative of affine connection is
 \begin{eqnarray}
 	L_\zeta \Gamma^{\lambda}{}_{\mu\nu}=-\nabla_\mu\nabla_\nu \zeta^{\lambda}+R_{\alpha\mu\nu}{}^{\lambda}\zeta^{\alpha}-\nabla_{\mu}(T^{\lambda}{}_{\alpha\nu}\zeta^{\alpha}).
 \end{eqnarray}
 Next we will consider that $\zeta^\mu$ is symmetric transformation, which means
 \begin{eqnarray}\label{2.lie gamma}
 	\mathcal{L}_{\zeta}g_{\mu\nu}=\mathring{\nabla}_{\mu}\zeta_{\nu}+\mathring{\nabla}_{\nu}\zeta_{\mu}=0.
 \end{eqnarray}
 Since in STG gravity, the general affine connection can be represents as
 \begin{eqnarray} \label{Gamma=Gamma+S}
 	{\Gamma^{\lambda}}_{\mu \nu}  =  {\mathring{\Gamma}^{\lambda}}_{\mu \nu} + {S^{\lambda}}_{\mu\nu},\quad 	S_{\lambda\mu\nu}= -\frac{1}{2}( Q_{\mu\nu\lambda} +Q_{\nu\mu\lambda} - Q_{\lambda\mu\nu} ).
 \end{eqnarray}
  Then we know that
  \begin{eqnarray}
  	\mathcal{L}_{\zeta}\Gamma^{\lambda}{}_{\mu\nu}=\mathcal{L}_{\zeta}\mathring{\Gamma}^{\lambda}{}_{\mu\nu}+\mathcal{L}_\zeta S^{\lambda}{}_{\mu\nu}.
  \end{eqnarray}
 Using cyclic symmetry of Riemann tensor $\mathring{R}_{[\mu\nu\rho]\sigma}=0$, we easily obtain
 \begin{eqnarray}
 	-\mathring{\nabla}_{\mu}\mathring{\nabla}_{\nu}\zeta^{\lambda}+\mathring{R}_{\alpha\mu\nu}{}^{\lambda}\zeta^{\alpha}=0.
 \end{eqnarray}
 Then using the Eq.(\ref{2.lie gamma}), we know that if metric satisfies symmetries of spacetime, then the Levi-Civita connection does as well. Consequently, the requirement that affine connection satisfies symmetries of spacetime is equivalent to the requirement that nonmetricity tensor satisfies the symmetries.
 
 Since the EOMs also contain covariant derivative of tensor, next we will prove that if affine connection and arbitrary tensor ($w_{\mu}$) satisfy the symmetries of spacetime, then the covariant derivative of $w_{\mu}$ does as well. Given that  $\nabla_\mu w_\nu$ is a tensor, then
 \begin{eqnarray}
 	\mathcal{L}_{\zeta}\nabla_\mu w_\nu=\mathcal{L}_{\zeta} (\mathring{\nabla}_{\mu}w_\nu - S^{\lambda}{}_{\mu\nu}w_\lambda)=\mathcal{L}_{\zeta} \mathring{\nabla}_{\mu}w_\nu -w_\lambda\mathcal{L}_{\zeta} S^{\lambda}{}_{\mu\nu}-S^{\lambda}{}_{\mu\nu}\mathcal{L}_{\zeta}w_\lambda.
 \end{eqnarray} 
 Since $\mathcal{L}_{\zeta} S^{\lambda}{}_{\mu\nu}=0$ and $\mathcal{L}_{\zeta}w_\lambda=0$, then we only need to prove
 \begin{eqnarray}
 	\mathcal{L}_{\zeta} \mathring{\nabla}_{\mu}w_\nu=0.
 \end{eqnarray}
 Further, using
 \begin{eqnarray}
 	\mathcal{L}_\zeta w_{\nu}=\zeta^\sigma \mathring{\nabla}_{\sigma}w_\nu+w_\sigma\mathring{\nabla}_\nu\zeta^\sigma=0, \quad \mathring{\nabla}_\mu (\mathcal{L}_\zeta w_{\nu})=0,
 \end{eqnarray}
 we finish the proof
 \begin{eqnarray}
 	\mathcal{L}_{\zeta} \mathring{\nabla}_{\mu}w_\nu&=&\zeta^\sigma \mathring{\nabla}_\sigma\mathring{\nabla}_\mu w_\nu+\mathring{\nabla}_\sigma w_\nu \mathring{\nabla}_\mu\zeta^\sigma +\mathring{\nabla}_\mu w_\sigma \mathring{\nabla}_\nu \zeta^\sigma\nonumber\\
 	&=&\zeta^\sigma \mathring{\nabla}_\sigma\mathring{\nabla}_\mu w_\nu-w_\sigma \mathring{\nabla}_\mu\mathring{\nabla}_\nu \zeta^\sigma-\zeta^\sigma \mathring{\nabla}_\mu\mathring{\nabla}_\sigma w_\nu\nonumber\\
 	&=&\mathring{R}_{\sigma\mu\nu\lambda}w^\lambda\zeta^\sigma-w_\sigma \mathring{\nabla}_\mu\mathring{\nabla}_\nu \zeta^\sigma=0.
 \end{eqnarray}
Finally we have proved that if affine connection also adheres to all symmetries of the background spacetime, the EOMs of gravitational part also respect these symmetries. From this perspective, it becomes evident why the off-diagonal terms of \(N^{\mu\nu}\) in Eq.(\ref{2.offdiagonal Nxx}) and Eq.(\ref{2.offdiagonal Npp}) do not vanish. When a affine connection complies with the symmetries of dS/AdS spacetime, we find that $N^{\mu\nu} = fg^{\mu\nu}$, where $f$ is a constant; however, this is not true for a zero affine connection.

\section{Perturbation analysis}\label{section3}

\subsection{Gauge transformation}

In this section, we will calculate the linear perturbations of action (\ref{2.action QQ}) on Minkowski spacetime and the quadratic action of perturbations is action (\ref{2.action HH}). To analyze the number of DOFs of this model, it is very convenient to use the “cosmological” decomposition in
terms of scalars, vectors and tensors under spatial rotations $SO(3)$  \cite{Alvarez:2006uu,Bahamonde:2024zkb,Mukhanov:1990me},
\begin{eqnarray}
	ds^2=-(1+2A)dt^2-2(\partial_iB+B_i)dtdx^i+\left[(1-2\psi)\delta_{ij}+2\partial_i\partial_jE+\partial_iE_j+\partial_jE_i+h_{ij}\right]dx^idx^j
\end{eqnarray}
and $u^{\mu}=\{u^0,\partial_iu+u_i\}$, where $A$, $B$, $\psi$, $E$, $u^{0}$, $u$ are six scalar perturbations, and $B_i$, $E_i$, $u_i$ are six vector perturbations which satisfy transverse condition, i.e., $\partial^iB_i=\partial^iE_i=\partial^iu_i=0$, and $h_{ij}$ are two tensor perturbations which are symmetric and satisfy the transverse and traceless conditions $\delta^{ij}h_{ij}=\partial^ih_{ij}=0$. Next we will consider the gauge transformation of these fundamental perturbation variables. Under the gauge transformation $\tilde{x}^{\mu}=x^{\mu}+\zeta^{\mu}$, we have the relations
\begin{eqnarray}
	\tilde{\delta g}_{\mu\nu}=\delta g_{\mu\nu}-\partial_{\mu}\zeta_{\nu}-\partial_{\nu}\zeta_{\mu},\quad \tilde{u}_{\mu}=u_{\mu}-\zeta_{\mu}.
\end{eqnarray}
So it is easy to see that $H_{\mu\nu}$ defined in Eq.(\ref{2.H=}) are gauge invariant variables. Using the decomposition  $\{\zeta^{\mu}={\zeta^{0},\zeta_i+\partial_i\zeta}\}$, the fundamental variables transform as
\begin{eqnarray}
	\tilde{A}=A-\partial_0\zeta^{0},\quad \tilde{B}=B+\partial_0\zeta-\zeta^{0},\quad \tilde{E}=E-\zeta,\quad \tilde{\psi}=\psi,\quad \tilde{u}^0=u^0-\zeta^0,\quad \tilde{u}=u-\zeta\nonumber\\
	\tilde{B}_{i}=B_i+\zeta'_i,\quad \tilde{E}_{i}=E_i-\zeta_i,\quad\tilde{u}_{i}=u_{i}-\zeta_i\nonumber\\
	\tilde{h}_{ij}=h_{ij}.
\end{eqnarray}
 Since $H_{\mu\nu}$ is gauge invariant and itself is a tensor, we can write it down using fully gauge invariant variables
\begin{eqnarray}
	H_{\mu\nu}=-2S_A(dt)_\mu(dt)_{\nu}-(\partial_iS_B+V^B_{i})(dt)_{\mu}(dx^i)\nu-(\partial_iS_B+V^B_{i})(dx^i)_{\mu}(dt)_{\nu}\nonumber\\+\left[-2S_\psi\delta_{ij}+2\partial_i\partial_jS_E+\partial_iV^E_{j}+\partial_jV^E_{i}+H^T_{ij}\right](dx^i)_\mu (dx^j)_\nu,
\end{eqnarray}
where we have 
\begin{eqnarray}
	S_{A}=A-\partial_0 u^0,\quad S_B=B+u'-u^0,\quad S_E=E-u,\quad S_{\psi}=\psi,\nonumber\\
	V^B_i=B_i+u'_i,\quad V^E_i=E_i-u_i,\nonumber\\
	H^T_{ij}=h_{ij}.
\end{eqnarray}
It is straightforward to verify that all of these variables are gauge invariant. In the following discussion, in order to simplify our calculations, we will adopt the so-called "coincident gauge," where $\zeta^0 = u^0$, $\zeta = u$, and $\zeta_i = u_i$. Consequently, the only remaining variables are the metric perturbations.

\subsection{Quadratic actions of perturbations}

In this subsection, we will separately calculate the quadratic actions for scalar, vector, and tensor perturbations. From these results, we can not only derive the linear perturbation equations but also determine the number of degrees of freedom (DOFs).

The quadratic action of tensor perturbations is
\begin{eqnarray}
	S^{(2)}_T=\frac{1}{2}\int d^4x c_1 (-\partial_t h_{ij}\partial_t h_{ij}+\partial_l h_{ij}\partial_l h_{ij}).
\end{eqnarray}
Tensor perturbations $h_{\mu\nu}$ represent the transverse and traceless gravitational waves (GWs). Since we are considering a gravity theory, we hope the models have GWs signals. Therefore, in this paper, we impose the condition that $c_1\neq 0$. From this quadratic action, it can also be inferred that the speed of GWs is equal to one (the speed of light). To prevent ghost instabilities, it is customary to require that  $c_1<0$.

The quadratic action of vector perturbations is
\begin{eqnarray}
	S^{(2)}_{V}=\frac{1}{2}\int d^4 x \left[(2c_1+c_2)\partial_tB_i\partial_tB_i-2\partial_t\partial_iE_j\partial_t\partial_iE_j+2c_2\partial_iB_j\partial_t\partial_iE_j-2c_1\partial_iB_j\partial_iB_j+(2c_1+c_2)\partial_i\partial_jE_k\partial_i\partial_jE_k\right].
\end{eqnarray}
In the following, we transform this action to Fourier space, where the Fourier transformation is
\begin{eqnarray}
	f(t,x)=\frac{1}{(2\pi)^{3/2}}\int d^3k f(t,k)e^{ikx}.
\end{eqnarray}
Since $B_i$ and $E_i$ are transverse vector, it is convenient to express them using circular polarization bases,
\begin{eqnarray}
	B_i=B^L e^L_i+B^R e^R_i,\quad E_i=E^L e^L_i+E^R e^R_i,
\end{eqnarray}
where the bases satisfy that $e^A_ie^{B*}_i=\delta^{AB}$ and $A,B=R,L$ represent the left and right-handed polarizations respectively. Then the quadratic action becomes
\begin{eqnarray}\label{3.action vector}
	S^{(2)}_V=\frac{1}{2}\sum_{A=L,R}\int dtd^3k \left[(2c_1+c_2)\partial_tB^A\partial_tB^A-2c_1k^2\partial_tE^A\partial_tE^A+2c_2k^2B^A\partial_tE^A-2c_1k^2B^AB^A+(2c_1+c_2)k^4E^AE^A\right],\nonumber\\
\end{eqnarray}
where have simply marked $B^AB^A*$, $E^AE^A*$ as $B^AB^A$, $E^AE^A$ and so on. It is easily seen that if $2c_1+c_2=0$, the perturbation variables $B^A$ are not dynamical fields and variation with respect to them will give us the constraint equation
\begin{eqnarray}
	B^A=\frac{c_2}{2c_1}\partial_t E^A.
\end{eqnarray}
Then taking it back into the action (\ref{3.action vector}), we find there are no propagating degrees of freedom of vector perturbations. This finding aligns with the results presented in \cite{Alvarez:2006uu}. This condition also corresponds to the so called transverse Fierz-Pauli symmetry or transverse diffeomorphisms (TDiff). If $2c_1+c_2\neq 0$, it is evidently that there are four DOFs for vector perturbations. The reference \cite{Alvarez:2006uu} has shown that if if there are no TDiff symmetries, the Hamiltonian of vector perturbations is not bounded below and generically this leads to a classical instability. In this paper, we will continue our analysis even in the case $2c_1+c_2\neq 0$. The reason are as follows. Firstly, the main goal of this paper is to establish a lower bound on the number of DOFs across various STG models. The action (\ref{2.action HH}) upon which we base our study is just the linear perturbation action, and we hope the nonlinear effects may mitigate this issue. Secondly, whether ghost modes is harmful or not still an open question, such as, recent research \cite{Deffayet:2021nnt,Deffayet:2023wdg,ErrastiDiez:2024hfq,Damour:2021fva} indicates that even when the Hamiltonian lacks a lower bound, stability may still be achievable within certain systems. Therefore, we also hope to explore alternative perspectives on addressing this problem.

The quadratic action for scalar perturbations in Fourier space is
\begin{eqnarray}\label{3.action scalar}
	S^{(2)}_S=\frac{1}{2}\int dtd^3k \left[-4(c_1+c_2+c_3+c_4)\partial_tA\partial_tA+12(2c_3+c_4)\partial_tA\partial_t\psi-12(c_1+3c_3)\partial_t\psi\partial_t\psi+(2c_1+c_2)k^2\partial_tB\partial_tB \right.\nonumber\\
	\left. +4(c_1+c_3)k^2A^2 -4(6c_3+c_4)k^2A\psi+4(3c_1+c_2+9c_3+3c_4)k^2\psi^2\right.\nonumber\\
	\left. -(2c_1+c_2)k^4B^2 -(4c_2+4c_4)k^2B\partial tA+4(c_2+3c_4)k^2B\partial_t\psi \right.\nonumber\\
	\left. -4(c_1+c_3)k^4\partial_t E\partial_t E-8(c_1+3c_3)k^2\partial_t\psi\partial_t E+4(2c_3+c_4)k^2\partial_tA\partial_t E+4(c_2+c_4)k^4 B \partial_t E\right.\nonumber\\
	\left. +4(c_1+c_2+c_3+c_4)k^6E^2-4(2c_3+c_4)k^4AE+8(c_1+c_2+3c_3+2c_`4)k^4\psi E  \right]\nonumber\\
\end{eqnarray}
For GR case, $c_1=-1/4$, $c_2=1/2$, $c_3=1/4$, $c_4=-1/2$, it reduces to
\begin{eqnarray}
	S^{(2)}_S=\frac{1}{2}\int dtd^3k \left[-6\partial_t\psi\partial_t\psi-4k^2A\psi-4k^2 B \partial_t\psi-4k^2 E'\partial_t\psi+2k^2\psi^2\right]
\end{eqnarray}
and since $A$, $B$ are all non-dynamical fields, they give the constraints
\begin{eqnarray}
	\psi=0,\quad \partial_t\psi=0.
\end{eqnarray}
Then we know that there are no propagating scalar perturbations in GR case. For other cases, it becomes evident that the presence or absence of TDiff symmetries ($2c_1+c_2$) significantly influences the dynamics of scalar perturbations, as this determines whether $B$ is a dynamical variable. Consequently, we will discuss these cases separately.

\subsubsection{Transverse diffeomorphisms}
For the case of transverse diffeomorphisms where $2c_1+c_2=0$, one can find that $B$ now is not a dynamical field and it gives us the constraint equation
\begin{eqnarray}
	(c_2+c_4)k^2E=(c_2+c_4)A-(c_2+3c_4)\psi.
\end{eqnarray}
Then if $c_2+c_4\neq 0$, replacing $E$ with $A$ and $\psi$ in the action (\ref{3.action scalar}), we can get the final form of quadratic action of scalar perturbations
\begin{eqnarray}\label{3.z2}
		S^{(2)}_S=\frac{1}{2}\int dtd^3k\, z^2 (\psi'(t)^2-k^2\psi^2), \quad with\quad z^2=\frac{4c_2(c_2^2-4c_2c_3+2c_2c_4+3c_4^2)}{(c_2+c_4)^2}.
\end{eqnarray}
From this quadratic action, we can obtain that if $c_2^2-4c_2c_3+2c_2c_4+3c_4^2\neq 0$, there are one scalar DOF; otherwise, there are no scalar DOF.

If $c_2+c_4=0$, we have $c_4\psi=0$. Since we require that $c_1\neq 0$, then $c_2=-c_4\neq 0$ and $\psi=0$. Then the final form of quadratic action of scalar perturbations is
\begin{eqnarray}
	S^{(2)}_S=\frac{1}{2}\int dtd^3k\, 2(c_2-2c_3)\left[(A-k^2E)^{'2}-k^2(A-k^2E)^2\right].
\end{eqnarray}
From this we know that if $c_2-2c_3\neq 0$, there are one scalar DOF; otherwise, there are no scalar DOF.

\subsubsection{No transverse diffeomorphisms}

For the case that $2c_1+c_2\neq 0$, the scalar perturbation $B$ has kinetic term, hence generally it is a dynamical field. The kinetic part of quadratic action for scalar perturbations can be written as
\begin{eqnarray}
	S^{(2)}_{Skin}=\frac{1}{2}\int dtd^3k\, \Phi^{T}K\Phi,
\end{eqnarray}
where we have define $\Phi=\{A,\psi,k^2E,kB\}$ and $K$ is the kinetic matrix
\begin{eqnarray}
	K=\begin{pmatrix}
		-4(c_1+c_2+c_3+c_4) & 6(2c_3+c_4) & 2(2c_3+c_4)&0\\
		6(2c_3+c_4)&-12(c_1+3c_3)&-4(c_1+3c_3)&0\\
		2(2c_3+c_4) &-4(c_1+3c_3)&-4(c_1+c_3)&0\\
		0&0&0&(2c_1+c_2)
	\end{pmatrix}.
\end{eqnarray}
The determinate of $K$ is
\begin{eqnarray}
	\det(K)=c_1(2c_1+c_2)\left[96(2c_3+c_4)^2-128(c_1+c_2+c_3+c_4)(c_1+3c_3)\right]
\end{eqnarray}
If $\det(K) \neq 0$, all four scalar perturbations are dynamical. Conversely, if $\det(K) = 0$, additional constraint equations arise, resulting in the number of degrees of freedom (DOFs) for scalar perturbations is less than four. In the case where $\det(K) = 0$, we focus on two specific cases: $c_3 + c_4 = 0$ and $c_1 + c_2 + c_3 + c_4 = 0$, as well as $c_3 + c_4 = 0$ and $c_1 + 3c_3$. The rationale for concentrating on these two cases is as follows. Reference \cite{Dambrosio:2020wbi} demonstrated that if the conditions $c_3 + c_4 = 0$ and $c_1 + c_2 + c_3 + c_4 = 0$ (with the constraints that $c_{1} \neq 0$ and $2c_{1} + c_{2} \neq 0$) are not met, there exist no primary constraints. Consequently, the number of DOFs associated with action (\ref{2.action QQ}) would be ten. Therefore, we only need to consider the case where both conditions—namely, $c_{3}+c_{4}=0$ and $c_{1}+c_{2}+c_{3}+c_{4}=0$ are satisfied. Additionally, we wish to examine the case where $ c_{3} + c_{4} = 0$ and $ c_{1} + 3c_{3}$, since it appears that in this situation, the number of DOFs at linear perturbation level is fewer than at nonlinear levels, indicating a potential strong coupling problem.

In the case $2c_3+c_4=0$, $c_1+c_2+c_3+c_4=0$, scalar perturbation $A$ is not a dynamical field, then it gives the constraint,
\begin{eqnarray}
	2(c_1+c_3)A=-(c_2+c_4)\partial_t B+(6c_3+c_4)\psi.
\end{eqnarray} 
If $c_1+c_3=0$, we have $c_2+c_4=0$, then the kinetic matrix of scalar perturbations becomes
\begin{eqnarray}
	K=\begin{pmatrix}
		0&0&0&0\\
		0&-12(c_1+3c_3)&-4(c_1+3c_3)&0\\
		0 &-4(c_1+3c_3)&0&0\\
		0&0&0&(2c_1+c_2)
	\end{pmatrix}.
\end{eqnarray}
From it, we can see that there are three DOFs. If $c_1+c_3\neq 0$, taking this constraint back into the action, we find that there are no kinetic term of $B$, then $B$ is now not a dynamical field and give us the constraint
\begin{eqnarray}
	(c_1+c_3)k^2B=-2(c_1+3c_3)\psi'-2(c_1+c_3)k^2E'.
\end{eqnarray}
Using this constraint, we obtain the final form of quadratic action of scalar perturbations
\begin{eqnarray}
	S^{(2)}_S=\frac{1}{2}\int dtd^3k\, \frac{-8c_1(c_1+3c_3)}{c_1+c_3}(\psi^{'2}-k^2\psi^2).
\end{eqnarray}
We can see that there are only one DOF.

In the case $2c_3+c_4=0$, $c_1+3c_3=0$, the perturbation variable $\psi$ is not a dynamical variable. Hence we have the constraint equation
\begin{eqnarray}
	2(3c_1+c_2+9c_3+3c_4)\psi=(c_2+3c_4)\partial_tB+(6c_3+c_4)A-2(c_1+c_2+3c_3+2c_4)k^2E.
\end{eqnarray}
Since $2c_1+c_2\neq 0$, $3c_1+c_2+9c_3+3c_4\neq 0$, we can replace $\psi$ with $B$, $A$ and $E$. Then we find that $B$ is also non-dynamical and give us the corresponding constraint
\begin{eqnarray}
	(2c_2-6c_3)k^2B=-2(c_2-4c_3)A'+4c_3k^2E'.
\end{eqnarray}
Using this constraint, we obtain the final form of quadratic action of scalar perturbations
\begin{eqnarray}
	S=\int dtd^3k\frac{8c_3(c_2-4c_3)}{2c_1+c_2}\left[(A'-k^2E')^2-k^2(A-k^2E)^2\right].
\end{eqnarray}
From this analysis, we conclude that the action (\ref{2.action QQ}) exhibits seven degrees of freedom (DOFs) on a Minkowski background at the linear perturbation level, under the condition $c_1\neq 0$, $2c_1+c_2\neq 0$, $2c_3+c_4=0$, $c_1+3c_3=0$.  However, reference \cite{Dambrosio:2020wbi} indicates that there are no primary constraints in this scenario; thus, the actual number of DOFs for the action (\ref{2.action QQ}) is ten. This implies that the model described by (\ref{2.action QQ}) may encounter strong coupling issues when considered in a flat background.

To summary this section, we give a table that shows that the lower bound of number of DOFs of the action (\ref{2.action QQ}).

\begin{table}[!h]
	\centering
	\begin{tabular}{c|c|c|c}
		\hline
		\multicolumn{3}{c|}{parameter}& The number of DOFs\\
		\hline
		\multirow{2}{*}{\makecell{$2c_1+c_2=0$}}&\multicolumn{2}{c|}{\makecell{$c_2+c_4=0, c_2-2c_3=0$} or \makecell{$c_2+c_4\neq 0, c_2^2-4c_2c_3+4c_2c_4+3c_4^2=0$}}&\makecell{$\geq 2$ }\\
		\cline{2-4}
		& \multicolumn{2}{c|}{otherwise}& \makecell{$\geq 3$ }\\
		\hline
		\multirow{3}{*}{\makecell{$2c_1+c_2\neq 0$}} &\multirow{2}{*}{\makecell{$2c_3+c_4=0, c_1+c_2+c_3+c_4=0$}}&\makecell{$c_1+c_3=0$}&$9$\\
		\cline{3-4}
		& & \makecell{$c_1+c_3\neq 0$}&$\geq 7$\\
		\cline{2-4}
		& \multicolumn{2}{c|}{otherwise} & \makecell{$10$}\\
		\hline
	\end{tabular}
	\caption[Table 1]{The lower bound of number of DOFs with $c_1\neq 0$}
\end{table}

\section{Couple to scalar field}\label{section4}

There are many STG models \cite{Jarv:2018bgs,Bahamonde:2022cmz} that incorporate a scalar field to formulate a scalar-tensor theory. Additionally, some STG models \cite{Bello-Morales:2024vqk,Zhao:2024kri} are constructed using a nonmetricity tensor but are dynamically equivalent to a scalar-tensor theory. For instance, the popular $f(Q)$ model is dynamical equivalent to a scalar-tensor theory  through the application of conformal transformation  \cite{Zhao:2024kri,Gakis:2019rdd}. Therefore, in this context, we will examine the scenario where gravitational interactions involve a scalar field.

The action of the scalar component that we will consider here is
\begin{eqnarray}\label{4.action phi}
	S_{\phi}=\frac{1}{2}\int d^4x \sqrt{-g} \left[b_1\partial^\mu \phi Q_\mu+b_2 \partial^\mu\phi \tilde{Q}_{\mu}+P(X,\phi)\right],
\end{eqnarray}
where $b_1$ and $b_2$ are constant, $X=\partial^\mu\phi\partial_\mu\phi$. This part can be considered as the action of matter field, then the corresponding energy momentum tensor is
\begin{eqnarray}
	T^{\mu\nu}_{\phi}=\frac{1}{2}g^{\mu\nu}\left(b_2\partial^\mu\phi\tilde{Q}_{\mu}+P\right)&&-b_1\left[\partial^{(\mu}\phi Q^{\nu)}-\tilde{Q}^\lambda \partial_\lambda \phi g^{\mu\nu}+\nabla^\lambda\nabla_\lambda\phi g^{\mu\nu}\right]\nonumber\\
	&&-b_2\left[\partial^\lambda\phi Q^{(\mu\nu)}{}_{\lambda}+\frac{1}{2}Q^{(\mu}\partial^{\nu)}\phi-Q^{(\mu\nu)\lambda}\partial_\lambda\phi+g^{\lambda(\mu}\nabla^{\nu)}\nabla_\lambda\phi\right]-P_X\partial^\mu\phi\partial^\nu\phi,
\end{eqnarray}
where $P_X\equiv \partial P/\partial X$, next we also will use notation $P_\phi\equiv \partial P/\partial \phi$. The EOM of scalar field is
\begin{eqnarray}
	-2\mathring{\nabla}_{\mu}(P_X\mathring{\nabla}^{\mu}\phi)-\mathring{\nabla}_{\mu}(b_1Q^\mu+b_2\tilde{Q}^{\mu})+P_\phi=0.
\end{eqnarray} 
If  there is a flat background (\ref{2.minkowski background}), the scalar field $\phi$ should satisfies the symmetries of this background, then $\phi=const$. On a flat background, form the EOMs, we have the relations that $P=0$ and $P_\phi=0$. Then for general case, such as $P=-X/2-m_{\phi}^2\phi^2/2$, we set $\phi=0$. Therefore $\phi$ itself is a perturbed quantity. 

Just like previous section, here we will calculate the quadratic actions of perturbations on flat background. It is easy to see that the scalar coupling term (\ref{4.action phi}) does not influence the tensor and vector perturbations. And the quadratic action for scalar perturbations of Eq.(\ref{4.action phi}) is
\begin{eqnarray}
	S^{(2)}_{\phi}=\frac{1}{2}\int dtd^3x\Big[&&-2(b_1+b_2)\partial_tA\partial_t\phi+6b_1\partial_t\psi\partial_t\phi+2b_1k^2\partial_tE\partial_t\phi-P_X\partial_t\phi\partial_t\phi\nonumber\\
	&&+2b_1k^2A\phi-2(3b_1+b_2)k^2\psi\phi-2(b_1+b_2)k^4E\phi-2b_2k^2B\partial_t\phi
	+(P_Xk^2+\frac{1}{2}P_{\phi\phi})\phi^2\Big].
\end{eqnarray}
To simplify our calculations, in this section, we only consider the TDiff case, i.e., $2c_1+c_2=0$. In this case, it is easily seen that $B$ is still not a dynamical variable, and we have the corresponding constraint
\begin{eqnarray}\label{4.B constraint}
	(c_2+c_4)k^2E=(c_2+c_4)A-(c_2+3c_4)\psi+\frac{1}{2}b_2\phi.
\end{eqnarray}
From this constraint equation, we can see that whether $b_2$ is zero or not will influence the constraint structure. Here we will consider the cases $b_2=0$ and $b_2\neq 0$ separately.

In the case which $b_2=0$, when $c_2+c_4=0$, we have $\psi=0$. Then using the constraint equations, the final quadratic action for scalar perturbations is
\begin{eqnarray}
	S^{(2)}_S=\frac{1}{2}\int dtd^3k\, 2(c_2-2c_3)\left[(A-k^2E)^{'2}-k^2(A-k^2E)^2\right]-2b_1(A-k^2E)'\phi'-P_X\phi^{'2}\nonumber\\
	+2b_1k^2(A-k^2E)\phi+(P_Xk^2+\frac{1}{2}P_{\phi\phi})\phi^2.
\end{eqnarray}
Then if the determent of kinetic matrix $2P_X(c_2-2c_3)-b_1^2$ is not zero, there are two scalar DOFs; otherwise, the number of DOFs is less than two. When $c_2+c_4\neq 0$, using the condition (\ref{4.B constraint}), the final quadratic action is
\begin{eqnarray}
	S^{(2)}_S=\frac{1}{2}\int dtd^3k\, z^2 (\psi^{'2}-k^2\psi^2)+\frac{4b_1c_2}{c_2+c_4}\phi'\psi'-P_{X}\phi^{'2}-\frac{4b_1c_2}{c_2+c_4}\phi\psi+(P_Xk^2+\frac{1}{2}P_{\phi\phi})\phi^2,
\end{eqnarray}
where $z^2$ have been defined in Eq.(\ref{3.z2}). Then if the main part of determent of kinetic matrix $c_2b_1^2+P_X(c_2^2-4c_2c_3+2c_2c_4+3c_4^2)$ is not zero, there are two scalar DOFs; otherwise, the number of DOFs of scalar perturbations is less than two.

In the case $b_2\neq 0$, when $c_2+c_4=0$, the constraint equation (\ref{4.B constraint}) becomes
\begin{eqnarray}
	4c_4\psi=b_2\phi
\end{eqnarray}
then the kinetic term of quadratic action of scalar perturbations is
\begin{eqnarray}
		S^{(2)}_{Skin}=\frac{1}{2}\int dtd^3k\, 2(c_2-2c_3)(A-k^2E)^{'2}-(2b_1-b_2+\frac{6b_2c_3}{c_2})(A-k^2E)'\phi'-(P_X+\frac{12b_1b_2c_2-3b_2^2c_2+18b_2^2c_3}{8c_2^2})\phi^{'2}.\nonumber\\
\end{eqnarray}
The determinant of kinetic matrix is proportional to
\begin{eqnarray}
	-b_1^2c_2-2b_1b_2c_2+\frac{1}{2}b_2^2(c_2-6c_3)-2c_2(c_2-2c_3)P_X.
\end{eqnarray}
Then if this determinant is not zero, there are two scalar DOFs; otherwise, the number of DOFs is less than two. When $c_2+c_4\neq 0$, the kinetic term of quadratic action of scalar perturbations is
\begin{eqnarray}
	S^{(2)}_{Skin}=\frac{1}{2}\int dtd^3k\,z^2\psi^{'2}+\frac{1}{2(c_2+c_4)^2}\Big\{8c_2\left[b_1(c_2+c_4)-b_2(2c_3+c_4)\right]\psi'\phi'\nonumber\\
	+\left[b_2^2(c_2-2c_3)+2b_1b_2(c_2+c_4)-2(c_2+c_4)^2P_X\right]\phi^{'2}\Big\}
\end{eqnarray}
The determinant of kinetic matrix is proportional to
\begin{eqnarray}
	2b_1^2c_2-b_2^2(c_2-6c_3)-2b_1b_2(c_2+3c_4)+2(c_2^2-4c_2c_3+2c_2c_4+3c_4^2)P_X
\end{eqnarray}
Then if this determinant is not zero, there are two scalar DOFs; otherwise, the number of DOFs is less than two.

From above discussion, we can summary this section. Even that we introduced the interaction term between nonmetricty tensor and scalar field, the additional scalar field generally only add one DOFs to the original action (\ref{2.action QQ}).

\section{Conclusion}\label{section5}

In this paper, we considered the most general STG action that is quadratic in nonmetricity tensor. We showed that Minkowski spacetime serves as a background solution. When adding a cosmological constant term to the action, we found that standard dS/AdS metrics with zero STG affine connection are not solutions. We also investigate non-vanishing affine connections which satisfy the symmetries of spacetime and frequently referenced in literature; however, these do not constitute solutions either. Thus, whether dS/AdS spacetime solutions exist in this model remains an open question. Furthermore, we provide proof that if an affine connection satisfies all symmetries of spacetime, then the equations of motion for the metric will similarly adhere to those symmetries. We believe this result may enhance our understanding of how to select appropriate affine connections within TG and STG theories.

Based on the Minkowski background solutions, we have calculated the quadratic actions for scalar, vector, and tensor perturbations. Utilizing these results, we have assessed the number of degrees of freedom associated with various parameter choices. Since this action represents the most general formulation that influences linear perturbations, our findings provide a lower bound STG models constructed using a nonmetricity tensor. Furthermore, leveraging previously published results regarding the number of primary constraints in this model allows us to establish an upper limit on the DOFs within STG theory. When considering the construction of STG models, it is common for researchers to incorporate scalar couplings. Accordingly, we also include potential scalar couplings that can arise at linear perturbation levels. For cases exhibiting transverse symmetry, we presented our calculations and counted the number of DOFs.

\subsection*{Acknowledgements}

 This work is supported by National Science Foundation of China (NSFC) under Grant No. 12347103, No. E414660101 and the Fundamental Research Funds for the Central Universities under Grants No. E3ER6601A2.

\bibliographystyle{utphys}
\providecommand{\href}[2]{#2}\begingroup\raggedright\endgroup

\end{document}